\documentclass[pra,twocolumn,showpacs,superscriptaddress,floatfix]{revtex4}
\usepackage{epsfig}
\usepackage{amsmath}

\newcounter{multieqn} 

\newcommand\br{\mathbf{r}}
\newcommand\rd{\mathrm{d}}

\begin{document}

\title{Force on a neutral atom near conducting microstructures}
\author{Claudia Eberlein}
\author{Robert Zietal}
\affiliation{Dept of Physics \& Astronomy,
    University of Sussex,
     Falmer, Brighton BN1 9QH, England}
\date{\today}
\begin{abstract}
We derive the non-retarded energy shift of a neutral atom for two different
geometries. For an atom close to a cylindrical wire we find an integral
representation for the energy shift, give asymptotic expressions, and
interpolate numerically. For an atom close to a semi-infinite halfplane we
determine the exact Green's function of the Laplace equation and use it
derive the exact energy shift for an arbitrary position of the atom. These
results can be used to estimate the energy shift of an atom close to etched
microstructures that protrude from substrates.
\end{abstract}


 \maketitle

\section{Introduction}
An important aim of current experimental cold atom physics is to learn how
to control and manipulate single or few atom in traps or along guides. To
this end more and more current and recent experiments deal with atoms close
to microstructures, most importantly wires and chips of various kinds
(cf. e.g. \cite{Schmiedmayr,Zimmermann,Hinds,Vuletic}). If these
microstructures carry strong currents then the resulting magnetic fields and
possibly additional external fields often create the dominant forces on the
atoms, which can then by used for trapping and manipulation. However, since
an atom is essentially a fluctuating dipole, polarization effects in those
microstructures lead to forces on the atoms even in the absence of currents
and external fields, and for atoms very close to them these can be
significant. Provided there is no direct wave-function overlap, i.e. the
atom is at least a few Bohr radii away from the microstructure, the only
relevant force is then the Casimir-Polder force \cite{CasimirPolder}, which
is the term commonly used for the van der Waals force between a point-like
polarizable particle and an extended object, in this case -- the atom and
the microstructure.

The type of microstructures that can be used for atom chips and can be
efficiently manufactured often involve a ledge protruding from an
electroplated and subsequently etched substrate \cite{private}. Here we
model this type of system in two ways: first by a cylindrical wire (which is
a good model for situations where the reflectivity of the electroplated top
layer far exceeds that of the substrate), and second by a semi-infinite
halfplane (which is an applicable model if the reflectivities of the top
layer and substrate do not differ by much). While in reality the
electromagnetic reflectivity of a material is of course never perfect, it
has been shown by earlier research that interaction with imperfectly
reflecting surfaces leads to a Casimir-Polder force that differs only by a
minor numerical factor from the one for a perfectly reflecting surface
\cite{wylie,wu}. Thus we shall consider only perfectly reflecting surfaces
here, which keeps the difficulty of the mathematics involved to a reasonable
level.

\section{Energy shift and Green's function}
We would like to consider an atom close to a reflecting surface and work out
the energy shift in the atom due to the presence of the surface. As is well
known, if the distance of the atom to the surface is much smaller than the
wavelength of a typical internal transition, then the interaction with the
surface is dominated by non-retarded electrostatic forces 
\cite{CasimirPolder,Barton,wu}.
At larger distances retardation matters, but at the same time the
Casimir-Polder force is then significantly smaller and thus hard to measure
experimentally (see e.g. \cite{sukenik,inguscio, cornell}). In this paper we
shall concentrate on small distances that lie within the non-retarded
electrostatic regime, as these lie well within the range of the
experimentally realizable distances of cold atoms from microstructures
\cite{private}.

In order to determine this electrostatic energy shift one
needs to solve the (classical) Poisson equation for the electrostatic
potential $\Phi$
\begin{equation}
-\pmb{\nabla}^2\Phi=\frac{\rho}{\varepsilon_0}
\label{poisson}
\end{equation}
with the boundary condition that $\Phi=0$ on the surface of a perfect
reflector. The charge density $\rho$ would be the atomic dipole $\pmb{\mu}$
of charges $\pm q$ separated by a vector ${\bf D}$, i.e.
\begin{equation}
\rho(\br)=\lim_{{\bf D}\rightarrow 0}\, q \left[\, 
\delta^{(3)}(\br-(\br_0+{\bf D})) - \delta^{(3)}(\br-\br_0) \right]
\label{rho}
\end{equation}
for a dipole located at $\br_0$. Since the dipole is made up of two point
charges, one can find a solution to the Poisson equation (\ref{poisson}) via
the Green's function $G(\br,\br')$, which satisfies
\begin{equation}
-\pmb{\nabla}^2 G(\br,\br')= \delta^{(3)}(\br-\br')\;,
\label{green}
\end{equation}
subject to the boundary condition that it vanishes for all points $\br$ that
lie on the perfectly reflecting surface. For the purposes of this paper it
is advantageous to split the Green's function in the following way,
\begin{equation}
G(\br,\br') = \frac{1}{4\pi |\br-\br'|} + G_H(\br,\br')\;.
\label{GH}
\end{equation}
The first term is the Green's function of the Poisson equation in unbounded
space, and thus the second term is a solution of the homogeneous Laplace
equation, chosen in such a way that the sum satisfies the boundary
conditions required of $G(\br,\br')$. 

The total energy of the charge distribution is given by \cite{stratton}
\begin{equation}
E=\frac{1}{2} \int\rd^3\br\; \rho(\br)\Phi(\br)\;,
\end{equation}
with
\begin{equation}
\Phi(\br) = \frac{1}{\varepsilon_0}\int\rd^3\br'\ G(\br,\br')\,\rho(\br')
\end{equation}
Thus for a dipole with the charge density (\ref{rho}) the total energy is
\begin{eqnarray}
E=&&\hspace*{-3mm}\lim_{{\bf D}\rightarrow 0} \bigg\{
\frac{1}{8\pi\varepsilon_0} \frac{q^2}{|\br+{\bf D}-(\br+{\bf D})|}
+\frac{1}{8\pi\varepsilon_0} \frac{q^2}{|\br-\br|}\nonumber\\ 
&&-\frac{1}{8\pi\varepsilon_0} \frac{q^2}{|\br+{\bf D}-\br|}
-\frac{1}{8\pi\varepsilon_0} \frac{q^2}{|\br-(\br+{\bf D})|}\nonumber\\
&&+\frac{q^2}{2\varepsilon_0} \left[ G_H(\br_0+{\bf D},\br_0+{\bf D}) 
  - G_H(\br_0+{\bf D},\br_0) \right]\nonumber\\
&&-\frac{q^2}{2\varepsilon_0} \left[ G_H(\br_0,\br_0+{\bf D}) 
  - G_H(\br_0,\br_0) \right]\bigg\}\;.
\end{eqnarray}
The first two terms in this expression are the divergent self-energies of
the two point charges. The third and fourth terms are the energy of the
dipole, which also diverges in the limit ${\bf D}\rightarrow 0$. None of
these are interesting for us because they are the same no matter where the
dipole is located. The energy shift due to the presence of the surface is
given by the remaining four terms, which all depend just on the homogeneous
solution $G_H(\br,\br')$. In the limit ${\bf D}\rightarrow 0$ these four
terms can be written as derivatives of $G_H$, and thus we obtain for the
energy shift of a dipole $\pmb{\mu}=q{\bf D}$ due to the presence of the
surface
\begin{equation}
\Delta E = \left. \frac{1}{2\varepsilon_0} (\pmb{\mu}\cdot\pmb{\nabla})
(\pmb{\mu}\cdot\pmb{\nabla'})\; G_H(\br,\br') 
\right|_{\br=\br_0,\br'=\br_0}\;,
\end{equation}
where $\br_0$ is the location of the dipole.

When applying this to an atom one also needs to take into account that for an
atom without permanent dipole moment the quantum-mechanical expectation
value of a product of two components of the dipole moment is diagonal,
\begin{equation}
\langle \mu_i \mu_j \rangle = \langle \mu_i^2 \rangle\: \delta_{ij}\;,
\end{equation}
in any orthogonal coordinate system. This implies that for an atom the
energy shift due to the presence of the surface reads
\begin{equation}
\Delta E = \left. \frac{1}{2\varepsilon_0} \sum_{i=1}^3
\langle \mu_i^2 \rangle \nabla_i \nabla_i'\; G_H(\br,\br') 
\right|_{\br=\br_0,\br'=\br_0}\;.
\label{shift}
\end{equation}
Thus the central task in working out the non-retarded energy shift of an
atom in the vicinity of a surface is to work out the electrostatic Green's
function of the boundary-value problem for the geometry of this surface. In
the next two sections we are going to do this for two different surfaces:
for a cylindrical wire of radius $R$ and for a semi-infinite halfplane. In
either case we assume that the surfaces are perfectly reflecting, which
enforces the electrostatic potential $\Phi$ to vanish there.

\section{Non-retarded energy shift near a wire}\label{sec:wire}
\begin{figure}[t]
\vspace*{-1mm}
\begin{center}
\centerline{\epsfig{figure=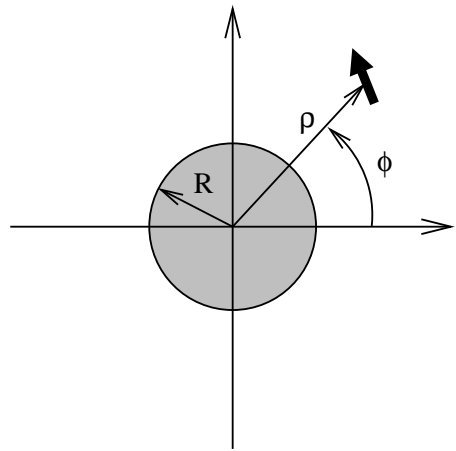,
  width= 6.0cm}}
\end{center}
\vspace*{-5mm} \caption{An illustration of the geometry of the dipole near a
wire. The radius of the wire is $R$, and the distance of the dipole from the
centre of the wire is $\rho$.}
\label{fig:wire}
\end{figure}
To calculate the energy shift, we first determine the Green's function of
the Poisson equation in the presence of a perfectly reflecting cylinder of
radius $R$ and infinite length. A standard method of calculating Green's
functions is via the eigenfunctions of the differential operator. In order 
to find a solution to Eq. (\ref{green}) we solve the eigenvalue problem
\begin{equation}
-\pmb{\nabla}^2\psi_n(\br)=\lambda_n\;\psi_n(\br)\;.
\end{equation}
The eigenfunctions $\psi_n(\br)$ must satisfy the same boundary conditions
as required of the Green's function. Since $-\pmb{\nabla}^2$ is Hermitean,
the set of all its normalized eigenfunctions must be complete,
\begin{equation}
\sum_n |\psi_n(\br)\rangle \langle\psi_n(\br')|=\delta^{(3)}(\br-\br')\;.
\end{equation}
Thus one can write
\begin{equation}
G(\br,\br')= \sum_n \frac{|\psi_n(\br)\rangle
  \langle\psi_n(\br')|}{\lambda_n}\;.
\label{eigensum}
\end{equation}
If we apply this method to unbounded space we can easily derive a
representation of the Green's function in unbounded space in cylindrical
coordinates,
\begin{eqnarray}
\frac{1}{4\pi|\br-\br'|}=\frac{1}{4\pi^2} \sum_{m=-\infty}^{\infty}
\int_{-\infty}^{\infty}\rd\kappa \int_0^\infty\rd k\ \frac{k}{k^2+\kappa^2}
\ \ \nonumber\\ \times
{\rm e}^{{\rm i}m(\phi-\phi')+{\rm i}\kappa(z-z')} J_m(k\rho) 
J_m(\kappa\rho')\;.
\label{freeG}
\end{eqnarray}
Performing the $k$ integration \cite[6.541(1.)]{GR}, we find
\begin{eqnarray}
\frac{1}{4\pi|\br-\br'|}=&&\hspace*{-3mm}
\frac{1}{2\pi^2} \sum_{m=-\infty}^{\infty}
\int_{0}^{\infty}\rd\kappa\ {\rm e}^{{\rm i}m(\phi-\phi')
+{\rm i}\kappa (z-z')}
\nonumber\\ &&\times
I_m(\kappa\rho) K_m(\kappa\rho')\;, \mbox{\ for\ } \rho<\rho'\;.
\label{GPC}
\end{eqnarray}
To derive the Green's function (\ref{GH}) that vanishes on the surface of
the cylinder $\rho=R$ we now just need to find the appropriate homogeneous
solution $G_H(\br,\br')$, which satisfies
\begin{equation}
-\pmb{\nabla}^2 G_H(\br,\br')=0\;.
\end{equation}
The general solution of the homogeneous Laplace equation in cylindrical
coordinates can be written as
\begin{equation}
\sum_{m=-\infty}^{\infty}
\int_{0}^{\infty}\rd\kappa\ {\rm e}^{{\rm i}m\phi+{\rm i}\kappa z}
\left[ A(m,\kappa) I_m(\kappa\rho) + B(m,\kappa) K_m(\kappa\rho) \right]
\end{equation}
where $A$ and $B$ are some constants.
Since $G(\br,\br')$ and therefore $G_H(\br,\br')$ must be regular at
infinite $\rho$ we must have $A(m,\kappa)=0$. This and the requirement that
the sum of Eq. (\ref{GPC}) and $G_H(\br,\br')$ must vanish at $\rho=R$ lead
to
\begin{eqnarray}
G_H(\br,\br')=&&\hspace*{-3mm}-\frac{1}{2\pi^2} \sum_{m=-\infty}^{\infty}
\int_{0}^{\infty}\rd\kappa\ {\rm e}^{{\rm i}m(\phi-\phi')
+{\rm i}\kappa(z-z')}\nonumber\\ &&\times
\frac{I_m(\kappa R)}{K_m(\kappa R)}\; K_m(\kappa\rho) K_m(\kappa\rho')\;.
\end{eqnarray}
The energy shift can now be determined by applying formula (\ref{shift}) in
cylindrical coordinates.
Taking into account the symmetry properties of the modified Bessel functions
we find for the energy shift of an atom whose location is given through
$(\rho,\phi,z)$ 
\begin{equation}
\Delta E = -\frac{1}{4\pi\varepsilon_0} \left[  
\Xi_\rho\langle\mu_\rho^2\rangle + \Xi_\phi\langle\mu_\phi^2\rangle
+ \Xi_z\langle\mu_z^2\rangle\right]
\label{dE}
\end{equation}
with the abbreviations
\begin{eqnarray}
\Xi_\rho&\!\!=&\!\! \frac{2}{\pi} \sum_{m=0}^{\infty}{}' 
\int_0^\infty\rd\kappa\ 
\kappa^2\frac{I_m(\kappa R)}{K_m(\kappa R)}\left[K_m'(\kappa\rho) \right]^2
\nonumber\\
\Xi_\phi&\!\!=&\!\! \frac{2}{\pi\rho^2} \sum_{m=1}^{\infty}\ m^2 
\int_0^\infty\rd\kappa\ 
\frac{I_m(\kappa R)}{K_m(\kappa R)}\left[K_m(\kappa\rho) \right]^2
\nonumber\\
\Xi_z&\!\!=&\!\! \frac{2}{\pi} \sum_{m=0}^{\infty}{}' \int_0^\infty\rd\kappa\ 
\kappa^2\frac{I_m(\kappa R)}{K_m(\kappa R)}\left[K_m(\kappa\rho) \right]^2\;.
\nonumber
\end{eqnarray}
The prime on the sums indicates that the $m=0$ term is weighted by an
additional factor 1/2. 

\begin{figure}
\vspace*{-1mm}
\begin{center}
\centerline{\epsfig{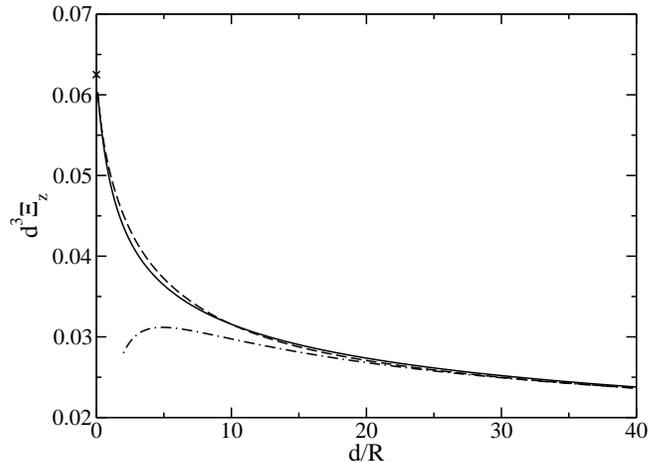}}
\end{center}
\vspace*{-5mm} \caption{The energy shift due to the $z$ component of the
  atomic dipole, multiplied by $d^3$, where $d$ is the distance of the dipole
  to the surface of the wire. The solid line is the exact expression
  calculated numerically, the dot-dashed line is the $m=0$ term alone, and
  the dashed line is the $m=0$ term plus the single integral derived through
  the uniform asymptotic approximation for the Bessel functions, 
  Eq. (\ref{approxZ}). The cross on the vertical axis gives the exact value 
  for $d\rightarrow 0$.}
\label{fig:XiZ}
\end{figure}

\begin{figure}
\vspace*{-1mm}
\begin{center}
\centerline{\epsfig{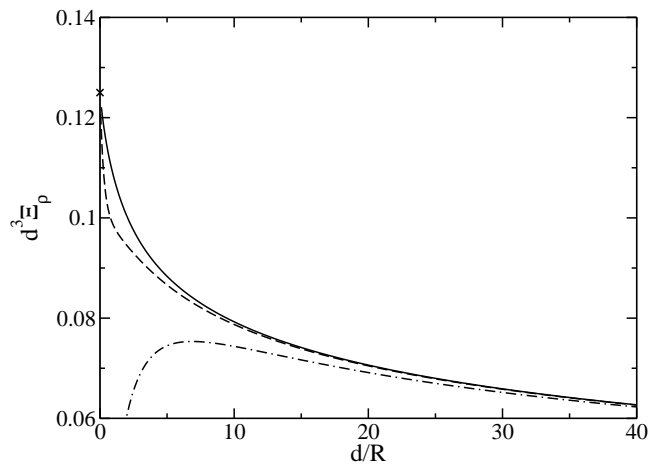}}
\end{center}
\vspace*{-5mm} \caption{Same as Fig. \ref{fig:XiZ} but for the energy shift
  due to the $\rho$ component of the atomic dipole. The solid line is the
  exact expression calculated numerically, the dot-dashed line is the $m=0$
  term alone, and the dashed line is the $m=0$ term plus the single integral
  derived through the uniform asymptotic approximation for the Bessel
  functions, Eq. (\ref{approxRho}).}
\label{fig:XiRho}
\end{figure}

\begin{figure}
\vspace*{-1mm}
\begin{center}
\centerline{\epsfig{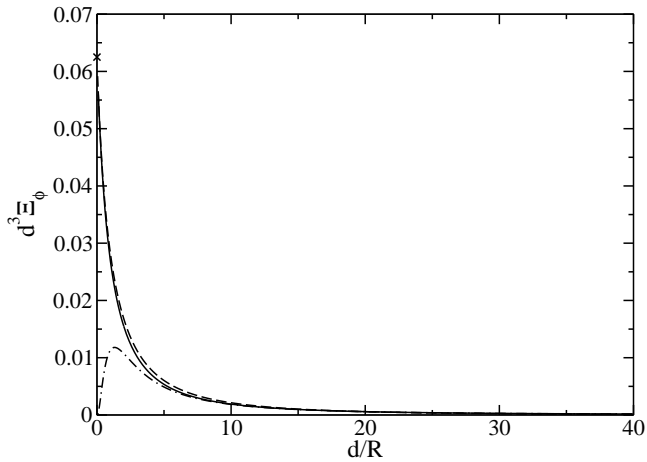}}
\end{center}
\vspace*{-5mm} \caption{Same as Fig. \ref{fig:XiZ} but for the energy shift
  due to the $\phi$ component of the atomic dipole.The solid line is the
  exact expression calculated numerically, the dot-dashed line is the $m=1$
  term alone, and the dashed line is the single integral
  derived through the uniform asymptotic approximation for the Bessel
  functions, Eq. (\ref{approxPhi}).}
\label{fig:XiPhi}
\end{figure}

Using numerical integration packages like those built into Mathematica or
Maple, one can evaluate these contributions to the energy shift. We show the
numerical results in Figs. \ref{fig:XiZ}--\ref{fig:XiPhi}. We have chosen to
show the various contributions to Eq. (\ref{dE}) as a function of the ratio
of the distance $d=\rho-R$ of the atom from the surface of the wire to the
wire radius $R$ and multiplied by $d^3$ so as to plot dimensionless quantities.

For most values of $d=\rho-R$ the integrals over $\kappa$ converge quite
well, as for large $\kappa$ the dominant behaviour of integrands is as
$\exp(-2\kappa d)$. Likewise is the convergence of the sums over $m$ very
good for reasonably large values of $d$. In fact, as shown by dot-dashed
lines in Figs. \ref{fig:XiZ}--\ref{fig:XiPhi}, convergence is so good that
from $d/R\approx 20$ upwards it is fully sufficient to take just the first
summand in each sum for $\Xi$.  However, convergence is less good for small
$d$ and thus the numerical evaluation of the energy shift gets more and more
cumbersome the closer the atom is to the surface of the wire. Neither the
integrals over $\kappa$ nor the sums over $m$ converge very well, so that it
is worthwhile finding a suitable approximation for small $d$. Another
motivation for a detailed analysis of the limit $d\rightarrow 0$ is of
course also to check consistency: if the atom is very close to the surface
of the wire ($d\ll R$) then the curvature of the wire cannot have any impact
on the shift any longer and the energy shift should simply be that of an
atom close to a plane surface, which is well known \cite{planeshift}.

To find a suitable approximation for small $d$, we separate off the $m=0$
terms in $\Xi_\rho$ and $\Xi_z$. In all the other summands and in $\Xi_\phi$
we scale by making a change of variables in the integrals to a new
integration variable $x=\kappa\rho/m$. The dominant contributions to those
integrals and the sums come from large $x$ and large $m$, so that one can
approximate the Bessel functions by their uniform asymptotic expansion
\cite[9.7.7--10]{AS}. Then the Bessel functions then the sums over $m$
become geometric series and can be summed analytically.  Taking just the
leading term in the uniform asymptotic expansions for the Bessel functions
we find the following approximations,
\renewcommand{\theequation}{\arabic{equation}%
\alph{multieqn}}\setcounter{multieqn}{1}%
\begin{eqnarray}
\Xi_\rho&\!\!\approx&\!\! \frac{1}{\pi}
\int_0^\infty\rd\kappa\ 
\kappa^2\frac{I_0(\kappa R)}{K_0(\kappa R)}\left[K_1(\kappa\rho) \right]^2
\nonumber\\&&\hspace*{7mm}
+\frac{1}{\pi\rho^3} \int_0^\infty\rd x\; 
\sqrt{1+x^2}\; \frac{A(A+1)}{(1-A)^3}
\label{approxRho}\\
\addtocounter{equation}{-1} \addtocounter{multieqn}{1}%
\Xi_\phi&\!\!\approx&\!\! \frac{1}{\pi\rho^3} \int_0^\infty\rd x\; 
\frac{1}{\sqrt{1+x^2}}\; \frac{A(A+1)}{(1-A)^3}
\label{approxPhi}\\
\addtocounter{equation}{-1} \addtocounter{multieqn}{1}%
\Xi_z&\!\!\approx&\!\! \frac{1}{\pi} \int_0^\infty\rd\kappa\ 
\kappa^2\frac{I_0(\kappa R)}{K_0(\kappa R)}\left[K_0(\kappa\rho) \right]^2
\nonumber\\&&\hspace*{7mm}
+\frac{1}{\pi\rho^3} \int_0^\infty\rd x \;
\frac{x^2}{\sqrt{1+x^2}}\; \frac{A(A+1)}{(1-A)^3}
\label{approxZ}
\end{eqnarray}
\renewcommand{\theequation}{\arabic{equation}}%
with the abbreviation
\[
A(x)=\frac{R^2}{\rho^2}\;{\rm e}^{-2\left(\sqrt{1+x^2}
-\sqrt{1+x^2\frac{R^2}{\rho^2}}
\right)}\left(\frac{1+\sqrt{1+x^2}}{1+\sqrt{1+x^2\frac{R^2}{\rho^2}}} 
\right)^2 .
\]
These are easy to evaluate numerically. We show the numerical values as
dashed lines in Figs. \ref{fig:XiZ}--\ref{fig:XiPhi}. Furthermore, these
approximations allow us to take the limit $d\rightarrow 0$. Taking this
limit under the $x$ integrals and retaining only the leading terms in each 
case, we can carry out the $x$ integrations analytically and obtain
\begin{equation}
\Xi_\rho \approx \frac{1}{8d^3}\ ,\ \ 
\Xi_\phi \approx \frac{1}{16d^3}\ ,\ \ 
\Xi_z \approx \frac{1}{16d^3}\;.
\label{Xi0}
\end{equation}
Inserting these values into Eq. (\ref{dE}) we see that they give the energy
shift of an atom in front a perfectly reflecting plane
\cite{planeshift}. This is an important consistency check for our
calculation, as the atom should not feel the curvature of the surface at
very close range. In Figs.  \ref{fig:XiZ}--\ref{fig:XiPhi} the limiting
values (\ref{Xi0}) are marked as crosses on the vertical axes. Since we have
taken along only first term of the uniform asymptotic expansion for each of
the Bessel functions, we cannot expect any agreement of the approximations
(\ref{approxRho}--c) beyond leading order. However, in practice these
approximations work quite well beyond leading order: as
Figs. \ref{fig:XiZ}--\ref{fig:XiPhi} show, approximations (\ref{approxPhi})
and (\ref{approxZ}) work very well for almost the entire range of $d$;
approximation (\ref{approxRho}) works reasonably well for small $d$ and then
again for large $d$ (because at large distances the $m=0$ term dominates
everything else). In this context we should also point out that the $m=0$
contributions to $\Xi_\rho$ and $\Xi_z$ do not contribute to leading order,
but behave as $d^{-2}$ in the limit $d\rightarrow 0$ and thus need to be
taken into account for asymptotic analysis in this limit beyond leading
order.

At large distances $d$ the energy shift (\ref{dE}) is dominated by the first
terms in each of the sums over $m$, i.e. the $m=0$ terms for $\Xi_\rho$ and
$\Xi_z$ and the $m=1$ term for $\Xi_\phi$. For large $d$ $\Xi_\rho$ and
$\Xi_z$  behave as
$1/(d^3 \ln d)$, but there is no point in giving an asymptotic expression as
further corrections are smaller only by additional powers of $1/\ln d$ and
these series converge far too slowly to be of any practical use. $\Xi_\phi$
falls off faster: to leading order we get $\Xi_\phi\approx 3\pi R^2/(32d^5)$
for large $d$.

Furthermore, if $d$ is not just large compared to $R$ but also compared to
the typical wavelength of an internal transition in the atom, then the
energy shift would be dominated by retardation effects, which cannot be
calculated with the electrostatic approach of this paper, but which require
a quantization of the electromagnetic field \cite{Barton}. For this reason
the extreme large-distance limit of the energy shift (\ref{dE}) is not of
practical importance and thus we see no reason to report on it in more
detail.

\section{Non-retarded energy shift near a semi-infinite halfplane}
\begin{figure}[t]
\vspace*{-1mm}
\begin{center}
\centerline{\epsfig{figure=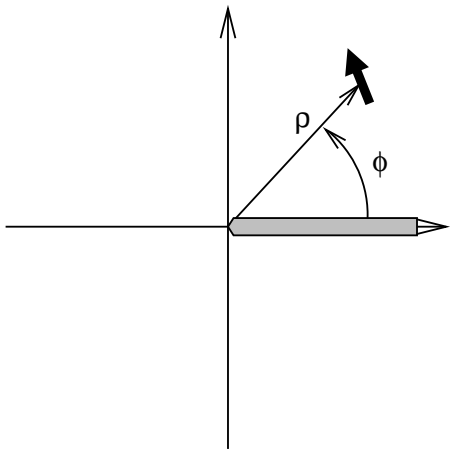,
  width= 6.0cm}}
\end{center}
\vspace*{-5mm} \caption{An illustration of the geometry of a dipole near a
  semi-infinite halfplane. The distance of the dipole from the edge is
  $\rho$, and from the surface of the plane it is $\rho\sin\phi$. }
\label{fig:halfplane}
\end{figure}

Next we wish to calculate the energy shift of an atom in the vicinity of a 
perfectly reflecting halfplane. The geometry is sketched in Fig. 
\ref{fig:halfplane}. 
The required boundary conditions are that the electrostatic potential
$\Phi(\br)$ vanishes at the angles $\phi=0$ and $\phi=2\pi$. We shall
calculate the Green's function for this case by using the method of
summation over eigenfunctions of the operator $-\pmb{\nabla}^2$, as
explained at the beginning of Section \ref{sec:wire}. If we determine the
set of eigenfunctions $\psi_n(\br)$ that satisfy the correct boundary
conditions then the Green's function (\ref{eigensum}) will also satisfy these
boundary conditions. Normalized eigenfunctions that vanish at $\phi=0$ and
$\phi=2\pi$ and that are regular at the origin and at infinity are in
cylindrical coordinates
\begin{equation}
\psi_n(\rho,\phi,z)=\frac{1}{\sqrt{2\pi}}\,{\rm e}^{{\rm i}\kappa z}\;
J_{m/2}(k\rho)\; \frac{1}{\sqrt{\pi}}\sin\frac{m\phi}{2}\;,
\end{equation}
with the corresponding eigenvalue $\lambda_n=\kappa^2+k^2$. Construction
(\ref{eigensum}) then gives the Green's function
\begin{eqnarray}
G(\br,\br')=\frac{1}{2\pi^2}\sum_{m=1}^{\infty}\int_0^\infty\rd k 
\int_{-\infty}^\infty\rd\kappa\ \frac{k}{\kappa^2+k^2}\;
{\rm e}^{{\rm i}\kappa(z-z')} \nonumber\\
\times\; J_{\frac{m}{2}}(k\rho)J_{\frac{m}{2}}(k\rho')\;
\sin\frac{m\phi}{2}\sin\frac{m\phi'}{2}\;.
\nonumber
\end{eqnarray}
This result is in agreement with the limiting case of the Green's function 
for the electrostatic potential at a perfectly reflecting wedge
\cite{smythe} if the wedge is to be taken to subtend a zero angle and extent
to an infinite radius.

As the energy shift (\ref{shift}) depends only on the homogeneous part
$G_H(\br,\br')$ of the solution, we need, according to Eq.~(\ref{GH}), to
subtract the free-space Green's function, which we have already written down
in Eq.~(\ref{freeG}). Noting that the free-space Green's function is of
course symmetric under the exchange of $\phi$ and $\phi'$, we can write it
in the form
\begin{eqnarray}
\frac{1}{4\pi|\br-\br'|}=\frac{1}{2\pi^2} \sum_{m=0}^{\infty}{}'
\int_{-\infty}^{\infty}\rd\kappa \int_0^\infty\rd k\ \frac{k}{k^2+\kappa^2}\;
{\rm e}^{{\rm i}\kappa(z-z')}\nonumber\\ \times
\cos m(\phi-\phi')\;J_m(k\rho) J_m(\kappa\rho')\;.
\nonumber
\end{eqnarray}
Taking the difference between $G(\br,\br')$ above and this expression and
carrying out the integration over $\kappa$ by closing the contour in the
complex plane and determining the residue, we obtain
\begin{widetext}
\begin{eqnarray}
G_H(\br,\br')&\!\!=&\!\!-\frac{1}{4\pi}\int_0^\infty\rd k\ 
{\rm e}^{-k|z-z'|} \left\{ \sum_{m=0}^{\infty}{}'\; J_m(k\rho)J_m(k\rho')
\left[ \cos m(\phi-\phi') + \cos m(\phi+\phi') \right] \right.\nonumber\\
&&\left.
+\sum_{m=0}^{\infty} J_{m+\frac{1}{2}}(k\rho)J_{m+\frac{1}{2}}(k\rho')
\left[ \cos \left(m+\frac12\right)(\phi+\phi') - \cos\left(m+\frac12\right)
(\phi-\phi') \right]\right\}\;.
\label{GHraw}
\end{eqnarray}
\end{widetext}
As before, primes on sums over $m$ indicate that the $m=0$ term is weighted
by an additional factor 1/2.

The sum in (\ref{GHraw}) over the product of Bessel functions with integer
indices can be carried out by applying standard formulae \cite[9.1.79]{AS},
\begin{eqnarray}
\sum_{m=0}^{\infty}{}'\; J_m(k\rho)J_m(k\rho') \cos m(\phi\pm\phi')
\hspace*{15mm}\nonumber\\
= \frac12 J_0(k\sqrt{\rho^2+\rho'^2-2\rho\rho'\cos(\phi\pm\phi')})\;.
\nonumber
\end{eqnarray}
Then the $k$ integration can also be carried out in these terms by applying
well known formulae \cite[11.4.39]{AS}. Thus we obtain for the terms
involving integer indices of the Bessel functions
\begin{eqnarray}
\int_0^\infty\rd k\ {\rm e}^{-k|z-z'|}
\sum_{m=0}^{\infty}{}'\; J_m(k\rho)J_m(k\rho') \cos m(\phi\pm\phi')
\hspace*{1mm}\nonumber\\
= \frac12 \frac{1}{\sqrt{(z-z')^2+\rho^2+\rho'^2
-2\rho\rho'\cos(\phi\pm\phi')}}\;.
\nonumber
\end{eqnarray}

The terms in (\ref{GHraw}) with Bessel functions of half-integer indices are
considerably more difficult to deal with. The only relevant formula we could
find anywhere is \cite[5.7.17.(11.)]{Prudnikov}
\begin{eqnarray}
\sum_{m=0}^{\infty} J_{m+\frac{1}{2}}(k\rho)J_{m+\frac{1}{2}}(k\rho')
\cos \left(m+\frac12\right)\alpha\hspace*{15mm}\nonumber\\
= \frac{1}{\pi}\int_{t_1}^{t_2}\rd t\ \frac{\sin t}{\sqrt{t^2-k^2\rho^2
-k^2\rho'^2+2k^2\rho\rho'\cos\alpha}}
\label{formula}
\end{eqnarray}
with the integration limits
\[
t_1=\sqrt{k^2\rho^2+k^2\rho'^2-2k^2\rho\rho'\cos\alpha}\ ,\ \ 
t_2=k(\rho+\rho')\;.
\]
Ref.~\cite{Prudnikov} does not give any references, so that the origin of
this formula cannot be traced. Although the ranges of applicability are
usually given for formulae in \cite{Prudnikov}, they are absent in this
particular case. However, inspection reveals that the formula cannot be
valid for the whole range $0\leq\alpha\leq 2\pi$ but must be restricted to
the range $0\leq\alpha\leq\pi$: the right-hand side has a periodicity of
$2\pi$, i.e.~it is the same for $\alpha=0$ and $\alpha=2\pi$, but the
left-hand side differs in sign for these two values of $\alpha$, as $\cos
0=1$ but $\cos (2m+1)\pi = -1$. Thus for the range
$0\leq|\phi\pm\phi'|\leq\pi$ we apply (\ref{formula}) with
$\alpha=\phi\pm\phi'$, and for the range $\pi\leq|\phi\pm\phi'|\leq 2\pi$ we
set $\alpha=2\pi-(\phi\pm\phi')$. If we scale the integration variable from
$t$ to $s=t/k$ then the subsequent $k$ integration is elementary,
\begin{equation}
\int_0^\infty\rd k\ {\rm e}^{-k|z-z'|}\sin ks = \frac{s}{s^2+(z-z')^2}\;.
\end{equation}
The remaining integration over $s$ can be carried out by changing variables
from $s$ to $v=s^2-s_1^2$, so that \cite[2.211]{GR}
\begin{eqnarray}
\int_{s_1}^{s_2}\rd s\ \frac{s}{s^2-(z-z')^2}\ \frac{1}{\sqrt{s^2-s_1^2}}
\hspace*{35mm}\nonumber\\
=\frac{1}{\sqrt{s_1^2+(z-z')^2}}\arctan
\sqrt{\frac{s_2^2-s_1^2}{s_1^2+(z-z')^2}}\;.\nonumber
\end{eqnarray}
Along these lines and distinguishing carefully between the cases
$0\leq|\phi\pm\phi'|\leq\pi$ and $\pi\leq|\phi\pm\phi'|\leq 2\pi$ we obtain
the following exact expression for the homogeneous part of the Green's
function
\begin{widetext}
\begin{eqnarray}
G_H(\br,\br')=-\frac{1}{4\pi} \left\{
\frac{1}{2\sqrt{(z-z')^2+\rho^2+\rho'^2-2\rho\rho'\cos(\phi-\phi')}}
+\frac{1}{2\sqrt{(z-z')^2+\rho^2+\rho'^2-2\rho\rho'\cos(\phi+\phi')}}
\right.\nonumber\\
+\frac{\mbox{sgn}(\sin|\phi+\phi'|)}{\pi
\sqrt{(z-z')^2+\rho^2+\rho'^2-2\rho\rho'\cos(\phi+\phi')}}
\;\arctan\sqrt{\frac{2\rho\rho'\left[1+\cos(\phi+\phi')\right]}{(z-z')^2+\rho^2
+\rho'^2-2\rho\rho'\cos(\phi+\phi')}}\nonumber\\
\left. -\frac{\mbox{sgn}(\sin|\phi-\phi'|)}{\pi
\sqrt{(z-z')^2+\rho^2+\rho'^2-2\rho\rho'\cos(\phi-\phi')}}
\;\arctan\sqrt{\frac{2\rho\rho'\left[1+\cos(\phi-\phi')\right]}{(z-z')^2+\rho^2
+\rho'^2-2\rho\rho'\cos(\phi-\phi')}}\right\}
\end{eqnarray}
\end{widetext}

Calculating the energy shift is now straightforward. We apply Eq.~(\ref{shift})
and find the exact energy shift of an atom located at $(\rho,\phi,z)$
\begin{equation}
\Delta E = -\frac{1}{4\pi\varepsilon_0} \left[  
\Xi_\rho\langle\mu_\rho^2\rangle + \Xi_\phi\langle\mu_\phi^2\rangle
+ \Xi_z\langle\mu_z^2\rangle\right]
\label{dE2}
\end{equation}
with the abbreviations
\begin{eqnarray}
\Xi_\rho&\!\!=&\!\!\frac{5}{48\pi\rho^3}+\frac{\cos\phi}{16\pi\rho^3\sin^2\phi}
+\frac{(\pi-\phi)(1+\sin^2\phi)}{16\pi\rho^3\sin^3\phi}\nonumber\\
\Xi_\phi&\!\!=&\!\!-\frac{1}{48\pi\rho^3}
+\frac{\cos\phi}{8\pi\rho^3\sin^2\phi}
+\frac{(\pi-\phi)(1+\cos^2\phi)}{16\pi\rho^3\sin^3\phi}\nonumber\\
\Xi_z&\!\!=&\!\!\frac{1}{24\pi\rho^3}+\frac{\cos\phi}{16\pi\rho^3\sin^2\phi}
+\frac{\pi-\phi}{16\pi\rho^3\sin^3\phi}\nonumber\;.
\end{eqnarray}
Here the applicable range of $\phi$ is $0\leq\phi\leq\pi$. Geometries with
$\phi$ in the range $\pi\leq\phi\leq 2\pi$ are obviously just mirror-images
of those with $0\leq\phi\leq\pi$, so that one could simply replace $\phi$ by
$2\pi-\phi$.

An important cross-check is the limit $\phi\rightarrow 0$. Very
close to the surface the edge of the half-plane should not affect the energy
shift, as the distance of the atom to the edge is $\rho$ but the distance to
the surface is $d\approx\rho\phi\ll\rho$. Thus the energy shift should be
the same as that in front of a plane. Indeed, the leading terms in the limit 
$\phi\rightarrow 0$ are
\begin{equation}
\Xi_\rho\approx\frac{1}{16\rho^3\phi^3}\ ,\ \
\Xi_\phi\approx\frac{1}{8\rho^3\phi^3}\ ,\ \ 
\Xi_z\approx\frac{1}{16\rho^3\phi^3}\;,
\end{equation}
which, if inserted into (\ref{dE2}), give the energy shift of an atom in
front of an infinitely extended reflective plane a distance $\rho\phi$ away.
Note that in contrast to Eq.~(\ref{Xi0}) for an atom very close to a
cylindrical surface, the component of the dipole that is normal to the
surface is now $\mu_\phi$.

\section{Summary}
We have calculated the non-retarded Casimir-Polder force on a neutral atom
that is either a distance $d$ away from a cylindrical wire of radius $R$, or
somewhere close to a semi-infinite halfplane. In both cases we have, for
simplicity, restricted ourselves to calculating the force for the atom
interacting with a perfectly reflecting surface. We have worked in
cylindrical coordinates $(\rho,\phi,z)$ and chosen the $z$ axis along the
centre of the wire and along the edge of the halfplane, respectively. The
energy shifts in a neutral atom due to the presence of the reflecting
surface nearby depend on the mean-square expectation values of the dipole
moments of the atom along the three orthogonal directions. For an atom close
to a reflecting wire the shift is given by Eq.~(\ref{dE}), and for an atom
near a semi-infinite halfplane by Eq.~(\ref{dE2}). The problem of the atom
and the wire has two independent length scales, the radius of the wire and
the distance of the atom from the wire, and accordingly the shift varies
with the ratio between them. We have provided analytical and numerical
approximations for both small and large values of this ratio in Section
III. By contrast, in the problem of an atom close to a semi-infinite
halfplane the distance of the atom to the surface is the only available
length scale, so that the energy shift does not depend on any further
parameters. This suggests that one could find an exact expression for the
energy shift, as we have indeed managed to do. It is nevertheless surprising
that this exact expression, Eq.~(\ref{dE2}), can be given in terms of
elementary functions and is so simple. Both model situations can be useful for
estimating the energy shift of an atom close to microstructures that consist
of a ledge and possibly an electroplated top layer.

\begin{acknowledgments}
It is a pleasure to thank Gabriel Barton for discussions. 
We would like to acknowledge financial support from The Nuffield Foundation.
\end{acknowledgments}

\end{document}